\documentclass[aps, prd, amsmath, amssymb, floats, floatfix, superscriptaddress, nofootinbib, twocolumn, showpacs,reprint]{revtex4}

\usepackage{graphicx}
\usepackage{amsmath}
\usepackage{amssymb}
\usepackage{amsfonts}
\usepackage{xspace} 
\usepackage[usenames]{color}
\usepackage{dcolumn}
\usepackage{bm}
\usepackage{mathrsfs}

\usepackage{verbatim}
\usepackage{appendix}

\usepackage{ulem}

\begin{document}

\title{Electromagnetic outflows in a class of scalar-tensor theories: binary neutron star coalescence}

\author{Marcelo Ponce}
\affiliation{Department of Physics, University of Guelph, Guelph, Ontario N1G 2W1, Canada.}

\author{Carlos Palenzuela}
\affiliation{Canadian Institute for Theoretical Astrophysics, Toronto, Ontario M5S 3H8, Canada.}
\affiliation{Departament de F\'isica, Universitat de les Illes Balears and Institut
d'Estudis Espacials de Catalunya, Cra. Valldemossa Km 7.5, E-07122 Palma de Mallorca, Spain.}

\author{Enrico Barausse}
\affiliation{CNRS, UMR 7095, Institut d'Astrophysique de Paris, 98bis Bd Arago, 75014 Paris, France.}
\affiliation{Sorbonne Universit\'es, UPMC Univ Paris 06, UMR 7095, 98bis Bd Arago, 75014 Paris, France.}

\author{Luis Lehner}
\affiliation{Perimeter Institute for Theoretical Physics, Waterloo, Ontario N2L 2Y5, Canada.}
\affiliation{CIFAR, Cosmology \& Gravity Program, Toronto, ON M5G 1Z8, Canada.}

\begin{abstract}
As we showed in previous work,  the dynamics and gravitational emission of binary neutron star systems in certain scalar-tensor theories can differ significantly
from that expected from General Relativity in the coalescing
stage. In this work we examine whether the characteristics of the electromagnetic counterparts to these binaries -- driven by
magnetosphere interactions prior to the merger event -- 
can provide an independent way to test gravity in the most strongly dynamical stages of binary mergers.
We find that the electromagnetic flux emitted by binaries in these scalar-tensor theories can show deviations from the GR prediction
in particular cases. These differences are quite subtle, thus requiring delicate 
measurements to differentiate between GR and the type of scalar-tensor theories considered in this work using electromagnetic observations alone. However, if coupled
with a gravitational-wave detection, electromagnetic measurements might provide a way to increase the confidence with which GR will be confirmed (or ruled out)
by gravitational observations. 
\end{abstract}


\date{\today}

\maketitle

\section{Introduction}
General Relativity (GR) has been very successful at describing gravity
in a vast range of scales, from submillimeter ones~\cite{Adelberger:2002ic}, to those of the Earth, Solar system~\cite{Will:2005va}
and binary pulsars~\cite{1975ApJ...195L..51H,1975ApJ...199L..25B,Damour:1998jk,Antoniadis:2013pzd,Freire:2012mg}.
This beautiful theory, however, is known to be incomplete in the ultra-violet regime where
it must be replaced by a quantum theory of gravity. Furthermore, at  
the infra-red cosmological scales, it needs to be supplemented with Dark Matter
and an unnaturally valued cosmological constant to explain observations, which can also be interpreted as a sign of failure.
These reasons
have spurred intense, and ongoing, efforts exploring how to describe gravity at 
both classical and quantum levels, the latter being one of the most
challenging enterprises of modern physics. 

Restricting to the classical regime --where both laboratory experiments and astrophysical observations can help constrain
possible alternatives-- a large number of putative theories have already been significantly constrained or ruled out altogether 
(see e.g. Refs.~\cite{Will:2005yc,1995werp.book.....W}). In the particular case of astrophysical observations, binaries
involving pulsars have proved especially well suited for these studies~\cite{1975ApJ...195L..51H,1975ApJ...199L..25B,Damour:1998jk,Antoniadis:2013pzd,Freire:2012mg}.
Indeed, exquisite electromagnetic observations of
the pulsar signal allow following the binary's orbital dynamics and comparing it with predictions
from GR and other theories. To date this task has necessarily involved binaries
at relatively large separations and, correspondingly, low orbital velocities ($v/c \ll 1)$. Binaries in such configurations, as a consequence,
do not fully explore possible discrepancies that might arise at relativistic velocities $v/c \simeq 1$.
Such discrepancies include dipolar emission of non-tensorial gravitational waves~\cite{eardley1975,Will:1989sk,Damour:1992we,Damour:1993hw,Damour:1996ke},
as well as dynamical violations of the (strong) equivalence principle
and enhancement of the strength of the gravitational attraction in the last, highly relativistic stages of the binary inspiral and plunge~\cite{Barausse:2012da,Shibata:2013pra,Palenzuela:2013hsa,Taniguchi:2014fqa}.

Such status of affairs will soon be radically changed thanks to a 
network of gravitational wave detectors that will allow analyzing compact binaries in highly relativistic velocity regimes. The detection
of gravitational waves from these systems will not only allow testing GR, but will additionally provide
important clues about the physical nature of these binaries, as well as identify the source's location. This knowledge may help -- both
directly and indirectly --  identify
electromagnetic counterparts (e.g. Refs.~\cite{2009arXiv0902.1527B,2013PhRvD..87l3004K,Andersson:2013mrx,Lehner:2014asa}) to these systems. 
The synergy of gravitational and electromagnetic observations will permit in-depth ``multi-messenger'' investigation
of the behavior of gravity in such highly relativistic binaries.
For this analysis to be possible, on the gravitational-wave side it is  
important that
possible deviations from the predictions of GR are either
understood, or suitably parametrized~\footnote{See discussion in e.g. Ref.~\cite{Sampson:2013jpa}.}, so as to guide
the detection and analysis of possible signals. Activities on both these fronts have recently been gaining significant momentum: 
In the ``parametrized'' approach particular formalisms have been presented, motivated by specific theories
and phenomenological considerations. Such formalisms have been applied to derive bounds on the relevant 
parameters describing deviations from GR~\cite{Yunes:2009ke,Li:2011cg,Loutrel:2014vja}.
In the ``direct'' approach, deviations from the gravitational waves predicted by  GR are computed in specific  bona fide gravity theories
(i.e. ones with a well-defined initial value problem)~\cite{Barausse:2012da,2012CQGra..29w2002H,Mirshekari:2013vb,Palenzuela:2013hsa,Shibata:2013pra,Berti:2013gfa}, and the prospects for
detecting these deviations are analyzed~\cite{Sampson:2013jpa,Sampson:2014qqa}. Among these theories, scalar-tensor
(ST) theories~\cite{Fierz:1956zz,jordan,Brans:1961sx,Bergmann:1968ve,Wagoner:1970vr,Nordtvedt:1970uv} -- where gravity is mediated not only by a metric tensor but also 
by a scalar field -- have received the most attention, because the presence of a scalar field in nature is  motivated 
by e.g., the low-energy limit of string theories, the observation of the Higgs boson, and cosmological phenomenology (i.e. inflation and Dark Energy).

Among the most likely sources of gravitational waves for Earth-based detectors is the coalescence of binary
neutron stars. The study of such systems in ST theories has been traditionally undertaken via suitable perturbation 
expansions~\cite{Will:1989sk,Damour:1992we,Damour:1996ke,Mirshekari:2013vb}, and more recently via numerical simulations~\cite{Barausse:2012da,Shibata:2013pra} or a hybrid 
approach~\cite{Palenzuela:2013hsa}. These works have not only provided definitive predictions for the expected signals
but have also helped illustrate that strong deviations from GR could arise, in certain classes of ST theories, because of dynamically-induced
effects as the orbit tightens~\cite{Barausse:2012da}. A first analysis indicating how such differences could be 
detected in the near future (by the upcoming second generation of gravitational wave interferometers) has been presented
in Ref.~\cite{Sampson:2014qqa} (see also~\cite{Taniguchi:2014fqa}). 
As mentioned, among the possible physical parameters that can be obtained via gravitational wave observations
are the time (and frequency) of the merger as well as the sky location, both of which would aid follow-up efforts
to capture counterparts in a wide range of electromagnetic bands. Within GR, much effort has been going into identifying promising near-coalescence scenarios 
 able to yield detectable signals in the electromagnetic spectrum,  and a number of mechanisms have been proposed and explored in recent years
(e.g. in Refs.~\cite{Palenzuela:2013kra,Palenzuela:2013hu,Kyutoku:2012fv,2012ApJ...746...48M,2012arXiv1204.6242P,Tsang:2011ad,2013PhRvD..87l3004K}).
It is therefore interesting to consider whether electromagnetic signals (either on their own, or in combination with gravitational wave observations) 
can provide additional clues as to whether gravity behaves as predicted by GR. This is especially important
as facilities for gravitational wave observations will be for years much more restricted in the frequency window they can access
in comparison to the large spectra provided by diverse electromagnetic observatories.
To this goal, we consider here whether electromagnetic 
counterparts triggered during the coalescence stage of neutron-star binaries can provide such testing opportunities. In particular
we here study the electromagnetic energy flux produced as a result of magnetosphere interactions in GR, as well as
in the ST theories that were shown in Refs.~\cite{Barausse:2012da,Palenzuela:2013hsa,Shibata:2013pra} to yield significant
deviations from the GR behavior in the late stages of the evolution of binary neutron star systems.

We take advantage of a unipolar induction model to account for these  magnetospheric effects, enhanced to include
the stars' own magnetization in defining the radius at which the induction takes place~\cite{Palenzuela:2013kra}.
The emitted electromagnetic Poynting luminosity  -- studied in Refs.~\cite{Ponce:2014sza,Palenzuela:2013kra,Palenzuela:2013hu} 
and related previous works~\cite{1969ApJ...156...59G,Hansen:2000am,Lai:2012qe} -- might be converted into potentially
observable x-ray and radio signals by several processes~\cite{Hansen:2000am,Palenzuela:2013kra}.
Additionally, the binary's dynamics needed for this model is described through a post-Newtonian (PN) treatment of the equations of
motions within scalar tensor theories, augmented by a set of equations that provide a description of the scalar charge of each binary's
component along the evolution~\cite{Palenzuela:2013hsa}.

This paper is organized as follows:
in section \ref{sec:Methods} we review the ST theories considered, and
we describe the procedure that we use to obtain the dynamics and estimate the electromagnetic luminosities;
in section \ref{sec:Results} we present the cases considered and the results obtained;
and finally in section \ref{sec:Discussion} we discuss the implications of our work.

\section{Methods}
\label{sec:Methods}

In this work we are primarily concerned with estimating possible
electromagnetic signals from the coalescence of magnetized binary neutron stars
in GR and certain ST theories. These theories are described in section~\ref{subsec:STdynamics}.
We here note that such theories do suffer from the infrared problems mentioned in our introduction
but do provide an interesting and well motivated model to explore possible ``ultraviolet'' (strong-field) deviations.
Since they also have a well defined initial value problem and yield a well posed physical problem,
even in the non-linear regime, they constitute an excellent arena to study possible deviations
from general relativity.
We obtain the binary's dynamics 
by solving the 2.5 PN equations of motion for ST theories as
described in Ref.~\cite{Mirshekari:2013vb} (enhanced with the formalism
proposed and validated in Ref.~\cite{Palenzuela:2013hsa} to account for the scalarization
effects allowed by the theories). The electromagnetic radiation 
induced by magnetospheric interactions of the magnetized neutron
stars is estimated using a phenomenological model
based on an extension of the unipolar inductor (we will refer to it as the ``unipolar model'' throughout this work). Such a model
captures magnetospheric effects by considering the electromotive force (emf)
that is induced as an otherwise non-magnetized conductor moves through a magnetic field~\cite{Hansen:2000am,Lai:2012qe}.
We have augmented this model recently in Ref.~\cite{Palenzuela:2013kra,Ponce:2014sza} to account
for an additional ``shielding effect'' that arises when both stars are magnetized, and which modifies
the radius at which the emf induction takes place.

\subsection{Scalar-tensor theories}
\label{subsec:STdynamics}

\subsubsection{Dynamics in ST theories and equations of motion}

The action for a ST theory can be written in the Jordan frame as
\begin{equation}
\label{Jframe_action}
S=\!\!\int d^4 x\frac{\sqrt{-g}}{2\kappa}\left[\phi R-\frac{\omega(\phi)}{\phi} \partial_\mu \phi \partial^\mu \phi
\right]+ S_M[g_{\mu\nu},\psi]\,,
\end{equation}
where  $\kappa=8 \pi G$, $R$, $g$ and $\phi$ are respectively the Ricci scalar, the metric determinant and the scalar field, and the
theory has no potential for the scalar field, but is characterized by an arbitrary function $\omega(\phi)$ (and by the boundary conditions for the scalar field).
Note also that
in this action we have assumed that the matter degrees of freedom $\psi$ couple minimally to the metric (and not to the
scalar field) so as to enforce the weak equivalence principle (i.e. the universality of free fall for weakly gravitating bodies).
The Jordan frame action can be recast in a more convenient form by a conformal transformation to the ``Einstein frame'', i.e. by defining a new metric $g_{\mu\nu}^E$ and a new scalar $\varphi$
such that $g^E_{\mu\nu}=\phi\, g_{\mu\nu}$ and  $({{\rm d}\log \phi}/{{\rm d}\varphi})^2={2\kappa}/[{3+2 \omega(\phi)}]$. This transformation casts the action \eqref{Jframe_action}
in the form 
\begin{equation}
\label{einframe}\!
\!S=\!\!\int\!d^4 x \sqrt{-g^E} \left( \frac{R^E}{2\kappa}\!-\!\frac{1}{2}g_E^{\mu\nu} \partial_\mu\varphi \partial_\nu\varphi
\right) \! +S_M\!\!\left[\frac{g^E_{\mu\nu}}{\phi(\varphi)},\psi\right]\!\!\!
\end{equation}
where note that the matter degrees of freedom still couple only to the Jordan frame metric  $g_{\mu\nu}=g^E_{\mu\nu}/\phi$ (i.e. test particles
follow geodesics of $g_{\mu\nu}$ and \textit{not} ones of $g^E_{\mu\nu}$), and that $g^E_{\mu\nu}$ and $\varphi$ are coupled minimally in the absence of matter
(which explains why using Einstein frame variables is advantageous).

By varying the Einstein frame action, one obtains the field equations
\begin{gather}\label{einstein}
G_{\mu\nu}^{E}=\kappa \left(T^\varphi_{\mu\nu} + T^E_{\mu\nu} \right),\\
\Box^E \varphi = \frac12 
\frac{{\rm d}\log \phi}{{\rm d}\varphi} T_E\label{KG} \,,\\
\nabla_\mu^E T_E^{\mu\nu}=-\frac{1}{2} T_E 
\frac{{\rm d}\log \phi}{{\rm d}\varphi} 
g_E^{\mu\nu} \partial_\mu \varphi\,,\label{scalarT}
\end{gather}
where we assume that indices are raised and lowered with $g^E_{\mu\nu}$ and define $T_E\equiv T_E^{\mu\nu}g^E_{\mu\nu}$.
Note that we also define the stress energy tensors appearing in the field equations as
\begin{eqnarray} 
T_E^{\mu\nu}&=&\frac{2}{\sqrt{-g^E}}\frac{\delta S_M}{\delta g^E_{\mu\nu}}=T^{\mu\nu}{\phi^{-3}} \, \\
T^\varphi_{\mu\nu} &=&\partial_\mu \varphi \partial_\nu \varphi- \frac{g^E_{\mu\nu}}{2} g_E^{\alpha\beta} \partial_\alpha \varphi
\partial_\beta \varphi\, \label{Tphi}
\end{eqnarray} 
where $T^{\mu\nu}$ is the Jordan frame stress-energy tensor of all the matter degrees of freedom.

Solutions to Eqs.~\eqref{einstein}--\eqref{scalarT} for binary systems of compact objects (e.g. neutron stars or black holes) can be obtained numerically in the late
stages of the inspiral and during the merger~\cite{Barausse:2012da,Shibata:2013pra,nico}, but in order to describe more widely separated systems such as observed binary pulsars, it is more convenient to expand the field equations in
 PN orders. In such a scheme, one approximates the two objects as point particles with masses $m_i$ and sensitivity parameters $s_i$~\cite{eardley1975} (or equivalently
 scalar charges~\cite{Damour:1992we} $\alpha_i= - (2 s_i-1)/(3+2\omega_0)^{1/2}$, with $\omega_0$ the value of the function $\omega(\phi)$ far from the binary system). The sensitivities can be calculated 
 from isolated solutions for the compact objects, and depend on the ST theory and on the object's compactness (e.g. $s_i\approx 0$ for white dwarfs and less compact stars, $s_i=1/2$ for black holes, while
 for neutron stars the sensitivity depends critically on the star's compactness and the ST theory under consideration). The binary's 
 dynamics is then expanded in orders of $v/c$ ($v$ being
 the binary's relative velocity), and to 2.5 PN order the resulting equations take the schematic form~\cite{Mirshekari:2013vb,Damour:1996ke,Damour:1992we,Will:1989sk}
\begin{eqnarray}
        \label{eqn:ST-2.5pn}
		&&\frac{d^2 \mathbf{x}}{dt^2} = -\frac{G_{\rm eff} M}{r^2} \mathbf{n}     \nonumber\\
                         &&+ \frac{G_{\rm eff} M}{r^2} \left[\left(\frac{\mathcal{A}_{PN}}{c^2}+\frac{\mathcal{A}_{2PN}}{c^4}\right)\mathbf{n}\
				+ \left( \frac{\mathcal{B}_{PN}}{c^2} + \frac{\mathcal{B}_{2PN}}{c^4} \right) \dot{r} \bf{v} \right] \nonumber\\
			&&	+ \frac{8}{5} \eta \frac{\left(G_{\rm eff} M\right)^2}{r^3} \left[\left( \frac{\mathcal{A}_{1.5PN}}{c^3} +\frac{ \mathcal{A}_{2.5PN}}{c^5}\right) \dot{r} \mathbf{n} \right.    \nonumber\\
			&&	\left.- \left( \frac{\mathcal{B}_{1.5PN}}{c^3} + \frac{\mathcal{B}_{2.5PN}}{c^5}\right) \mathbf{v}\right]
\end{eqnarray}
where $\mathbf{x} = \mathbf{x}_1 - \mathbf{x}_2$ is the binary separation,
$r = |\mathbf{x}|$, $\mathbf{n} = \mathbf{x}/r$,
$\mathbf{v}=\mathbf{v}_1 - \mathbf{v}_2$ is the relative velocity, $\dot{r} = dr/dt$, $M=m_1+m_2$ is the total mass of the system, and
$\eta = (m_1 m_2)/M^2$ is the symmetric mass ratio.
The ``effective'' gravitational constant $G_{\rm eff}$ is
related to the gravitational constant $G_N$ measured locally (e.g. by a Cavendish-type experiment) by
$G_{\rm eff} \approx G_N ( 1+\alpha_1\alpha_2 )$.
Explicit expressions for the functions $\{\mathcal{A_I},\mathcal{B_I}\}$ are given
in Ref.~\cite{Mirshekari:2013vb} and also depend on the
sensitivities/scalar charges of the binary components, e.g. the presence of dissipative 1.5PN terms (which are
absent in GR) in Eq.~\eqref{eqn:ST-2.5pn} is due to the scalar charges, which source the emission of
dipolar gravitational radiation with energy flux
\begin{equation}\label{dipole_flux}
\dot{E}_{\rm dipole} \approx \frac{G_N}{3 c^3} \left(\frac{G_{\rm eff} m_1 m_2}{r^2}\right)^2 (\alpha_1-\alpha_2)^2\,.
\end{equation}

\subsubsection{Spontaneous, induced and dynamical scalarization}
From the dependence of the PN equations on the sensitivities/scalar charges, which in turn depend on the nature and compactness of the binary's components, it is clear that
the PN evolution of a compact-object binary depends on the nature of its components. Therefore, the strong equivalence principle, defined as the universality of free fall 
for strongly-gravitating objects, is violated in ST theories already in the PN inspiral. Recently, however, Ref.~\cite{Barausse:2012da} highlighted the existence of other violations of the strong
equivalence principle in the last stages of the inspiral of binary neutron stars and for a particular class of ST theories. More specifically, Ref.~\cite{Barausse:2012da} considered theories with 
$\omega(\phi)=-3/2-\kappa/(4\beta \log\phi)$ (or equivalently $\phi = \exp(-\beta \varphi^2)$), which are known~\cite{Damour:1992we,Damour:1993hw} to give rise
to the ``spontaneous scalarization'' of isolated neutron stars, i.e. allow for scalar charges $\alpha\sim 0.1-1$ for sufficiently
compact neutron stars and for marginally viable values of the theory's parameters $\tilde{\beta}=\beta/(4\pi G)\gtrsim -4.5$ and $\varphi_0\lesssim 10^{-2}$ ($\varphi_0$ being the scalar's value far from the system).\footnote{We stress that the observational constraint $\tilde{\beta}\gtrsim -4.5$ depends
somewhat on the equation of state of neutron stars. Indeed, binary-pulsar observations essentially rule out spontaneous scalarization
for the neutron-star masses corresponding to the components of observed binaries, placing constraints on $\tilde{\beta}$ \textit{once} an equation of state is assumed. 
For instance, Ref.~\cite{Shibata:2013pra} shows that values of $\tilde{\beta}$ as low as $-5$ may be allowed with certain equations of state.
{Here, we follow Refs.~\cite{Barausse:2012da} and~\cite{Palenzuela:2013hsa}, and adopt
a polytropic equation of state with $K=123 G^3 M_\odot^2/c^6$
and $\Gamma=2$. Since this equation of state is not realistic but just a simple toy model, our results should be
interpreted as qualitative.}}
Ref.~\cite{Barausse:2012da} showed that at sufficiently small binary separations, a spontaneously scalarized star (thus bearing a significant scalar charge) 
can excite a scalar charge in the other star, even if that star had not spontaneously scalarized and thus did not have a scalar charge to start with. This ``induced scalarization''
was shown to be capable of triggering earlier binary plunges and mergers relative to GR, an effect potentially observable with advanced GW detectors~\cite{Sampson:2014qqa}. Even more strikingly, Ref.~\cite{Barausse:2012da} showed that significant
scalar charges can be produced in the last inspiral stages of binary systems of unscalarized neutron stars, i.e. in binaries whose components have little or no scalar charges at large separations.
This ``dynamical scalarization'' produces a sudden build-up of the scalar charges at small separations, quickly triggering a plunge/merger at frequencies within the reach of advanced GW detectors (c.f. Ref.~\cite{Sampson:2014qqa,Taniguchi:2014fqa}
for the detectability of this effect with GW detectors). It is also important to note that dynamical scalarization, being 
an effect that turns on at small separations, can evade (to a certain extent) the constraints posed
by binary-pulsar observations, which only exclude the presence of \textit{spontaneously} scalarized stars in widely-separated binaries and for the specific observed values
of the neutron-star masses. In particular, dynamical scalarization may happen for values of $\tilde{\beta}$ between $-4.3$ and $-4.5$ (or lower), a window still allowed by binary-pulsar observations.

Remarkably, even though both induced and dynamical scalarization are strongly non-linear effects, they can still be understood and reproduced in their main qualitative features by a minimal modification of the PN expansion
scheme outline above. More precisely, Ref.~\cite{Palenzuela:2013hsa} describes the evolution of a neutron-star binary system by the PN equations of motion \eqref{eqn:ST-2.5pn}, but introduces a new way of computing
the sensitivities or scalar charges that appear in those equations. Instead of computing those parameters from isolated neutron-star solutions, as done in the ``classic'' PN scheme, Ref.~\cite{Palenzuela:2013hsa} introduced a 
formalism that includes the interaction between the two stars in the calculation of the scalar charges, thus accounting for both induced and dynamical scalarization. In practice, this scheme starts from the
standard calculation of the scalar charges for the binary components
 in isolation, and then uses that calculation to define a system of non-linear algebraic equations, which can be solved iteratively at each step of the PN orbital evolution to yield the charges of 
 both stars including non-linear effects. This procedure was validated by 
 comparing it with the fully non-linear simulations of Ref.~\cite{Barausse:2012da}.

\subsubsection{Electromagnetic coupling in ST theories}
One purpose of this paper is to extend the formalism of Ref.~\cite{Palenzuela:2013hsa} to include the effect of an electromagnetic field.
Let us consider a binary system of magnetized neutron stars surrounded by a plasma in ST theories. Solving the full non-linear problem in the Jordan frame would require solving the curved spacetime Maxwell
equations
\begin{gather}
\partial_{[\mu} F_{\nu\gamma]}=0\,,\label{max1}\\
\nabla_\nu F^{\nu \mu} = j^\mu\,,\label{max2}
\end{gather}
and including the stress-energy tensor of the electromagnetic field in the source of the Einstein equations. (Note that because of the weak equivalence principle, which is reflected 
in the structure of the matter action written in Eq.~\eqref{Jframe_action}, the electromagnetic field only couples to the Jordan frame metric and to the plasma's electric charges, and not directly to the scalar field.)
Defining the Einstein-frame electromagnetic tensor $F^E_{\mu\nu}=F_{\mu\nu}$ and the plasma's current $j^\mu_E=j^\mu/\phi^2$, the Einstein-frame Maxwell equations take the same form as in the Jordan frame, i.e.
\begin{gather}
\partial_{[\mu} F^E_{\nu\gamma]}=0\label{m1}\,,\\
\nabla^E_\nu F_E^{\nu \mu} = j_E^\mu\,,\label{m2}
\end{gather}
and the Einstein-frame field equations \eqref{einstein}--\eqref{scalarT} become
\begin{gather}\label{einsteinBis}
G_{\mu\nu}^{E}=\kappa \left(T^\varphi_{\mu\nu}+ T^{{\rm em},E}_{\mu\nu}+ T^{{\rm pl},E}_{\mu\nu}+ T^E_{\mu\nu} \right),\\
\Box^E \varphi = \frac12 
\frac{{\rm d}\log \phi}{{\rm d}\varphi} (T_E+T_E^{\rm pl})\label{KGbis} \,,\\
\nabla_\mu^E (T_E^{\mu\nu}+T_{{\rm pl},E}^{\mu\nu}) =F^\nu_{E\mu}j_E^\mu-\frac{1}{2} (T_E+T_E^{\rm pl}) 
\frac{{\rm d}\log \phi}{{\rm d}\varphi} 
g_E^{\mu\nu} \partial_\mu \varphi\,,\label{scalarTbis}
\end{gather}
where $T_{{\rm em},E}^{\mu\nu}=T_{\rm em}^{\mu\nu}{\phi^{-3}}$ is the Einstein-frame stress-energy tensor of the electromagnetic field~\footnote{We recall that the explicit form of the stress-energy tensor of the electromagnetic
field in the Jordan frame is 
\begin{equation*}
T_{\rm em}^{\mu\nu}=\frac{1}{4 \pi} \left(F^{\mu\alpha} F^{\nu}_{\phantom{a}\alpha}-\frac14 g^{\mu\nu} F_{\alpha\beta}F^{\alpha\beta}\right)\,.
\end{equation*}
Similarly, the Einstein-frame stress energy tensor is given by
\begin{equation*}
T_{\rm em,\,E}^{\mu\nu}=\frac{1}{4 \pi} \left(F_E^{\mu\alpha} F^{\nu}_{E\,\alpha}-\frac14 g_E^{\mu\nu} F^E_{\alpha\beta}F_E^{\alpha\beta}\right)\,,
\end{equation*}
which satisfies indeed  $T_{{\rm em},E}^{\mu\nu}=T_{\rm em}^{\mu\nu}{\phi^{-3}}$.}, 
while $T_{E}^{\mu\nu}$,  $T_{{\rm pl},E}^{\mu\nu}$
and $T_{{\rm em},E}^{\mu\nu}$ are those of
the neutron-star matter, of the matter of the plasma and of the electromagnetic field.
(Again, all indices are raised and lowered with the Einstein-frame metric.)
As discussed below, the stress energy of the plasma matter will be negligible
for our cases of interest, but for the moment
we keep it to make our treatment general and more clear.
Note that in deriving these equations we have used
\begin{equation}\label{emCons}
\nabla_\mu^E T_{{\rm em},E}^{\mu\nu}=-F^\nu_{E\mu}j_E^\mu 
\end{equation}
(which follows from the Maxwell equations), and the fact that the trace of $T_{{\rm em},E}^{\mu\nu}$ is
zero. Note also that Eq.~\eqref{KGbis} can be recast in the form 
\begin{equation}
 \nabla^E_\mu T_\varphi^{\mu\nu}=  \frac12 
\frac{{\rm d}\log \phi}{{\rm d}\varphi}  (T_E^{\rm pl}+T_E) g_E^{\mu\nu} \partial_\mu \varphi\,,
 \end{equation}
which shows that in the Einstein frame (as in the Jordan frame) there is no direct energy or momentum transfer from the scalar field to the electromagnetic field.

Postponing the solution to the full non-linear problem to future work, let us note that for astrophysically realistic systems, the plasma and the electromagnetic field
in the magnetosphere are too small to significantly backreact on the metric and on the binary and scalar field evolution (``force-free approximation'', c.f. discussion in the next section), 
i.e. to lowest order Eqs.~\eqref{einsteinBis}--\eqref{scalarTbis} reduce to Eqs.~\eqref{einstein}--\eqref{scalarT}. To next order, by combining Eqs.~\eqref{scalarTbis} and \eqref{scalarT} one then obtains
\begin{equation}\label{plasmaMotion}
\nabla_\mu^E T_{{\rm pl},E}^{\mu\nu} =F^\nu_{E\mu}j_E^\mu-\frac{1}{2} T_E^{\rm pl}
\frac{{\rm d}\log \phi}{{\rm d}\varphi}g_E^{\mu\nu} \partial_\mu \varphi \approx F^\nu_{E\mu}j_E^\mu
\end{equation}
where we have used the fact that $T_E^{\rm pl}\approx 0$ for the plasma (because the particles of which it is made typically move at speeds close to the speed of light)
and in any case ${{\rm d}\log \phi}/{{\rm d}\varphi}\approx0$ outside the neutron stars (where the plasma moves) because $\varphi$ is small there. Equation~\eqref{plasmaMotion} regulates the motion
of the plasma, while the electromagnetic field satisfies the Maxwell equations \eqref{m1}--\eqref{m2}. In both Eq.~\eqref{plasmaMotion} and Eqs.~\eqref{m1}--\eqref{m2}, 
the metric is determined by the evolution of the binary and scalar field alone [i.e.~by Eqs.~\eqref{einstein}--\eqref{scalarT}].
In practice, as can be seen from Eq.~\eqref{Tphi},  the scalar field stress-energy vanishes at linear order in the field's perturbation over a constant background, and is thus negligible
outside the neutron stars (although it is \text{not} always negligible inside the stars, where it can grow non-linear and give rise to scalar charges $\alpha\sim 0.1-1$ in scalarized systems).
Therefore, it is natural to approximate the metric outside the neutron stars with the \textit{general-relativistic} PN metric of two point particles (representing the neutron stars), 
whose trajectories are calculated (\textit{including} the effect of the scalar charges) with the formalism of Ref.~\cite{Palenzuela:2013hsa}. 
Therefore, because of Eqs.~\eqref{m1}, \eqref{m2} and \eqref{plasmaMotion}, the calculation of the electromagnetic fluxes can proceed as in GR, except for the modified binary trajectory. We will present a standard GR approximate strategy to
calculate such fluxes (the ``unipolar inductor'' model) in the next section.

\subsection{Magnetosphere and plasma treatment}
\label{sec:magnetospherePlasma}
As discussed in Ref.~\cite{1969ApJ...157..869G}, neutron stars are surrounded by a magnetosphere with
a plasma density $\rho^{pl} \simeq -\vec{\Omega}\cdot\vec{B}/(2\pi c)$, where $\vec{\Omega}$ represents the
rotational frequency of the plasma, $\vec{B}$ is the magnetic field present in the region, and $c$ is the speed of light.
The interaction of a rotating magnetized star with
its own magnetosphere is responsible for electromagnetic emissions in pulsars. The analysis of such an interaction
is a delicate subject, because the plasma dynamics may be intricate, and complex simulations are typically required to
fully capture its behavior. Fortunately, a useful approximation can be adopted that captures important
aspects of the system. This approximation relies on the observation that in the magnetosphere region 
the inertia of the plasma is negligible with respect to the electromagnetic energy density, i.e. $T_{\mu\nu}^{pl} \ll T_{\mu\nu}^{em}$.
Through Eqs.~(\ref{emCons}) and~(\ref{plasmaMotion}), this in turn implies
$F^{\nu}_{E\mu} j^{\mu}_E\approx F^{\nu}_{\phantom{E}\mu} j^{\mu}\approx0\approx\nabla_{\mu}^{E} T_{em,E}^{\mu \nu}\approx\nabla_{\mu} T_{em}^{\mu \nu}$,
where in the last passage we have exploited the fact that Eq.~(\ref{emCons}) also holds in the Jordan frame (as it follows directly from the Maxwell equations~\eqref{max1} and \eqref{max2}).\footnote{One
can also show directly that $\nabla^E_\mu T^{\mu}_{em,E\nu}=\nabla_\mu T^{\mu}_{em\,\nu}/\phi^2$, using the conformal transformation between the Einstein and Jordan frames.}
These are known as the force-free conditions for the plasma~\cite{1969ApJ...156...59G,1977MNRAS.179..433B,2002MNRAS.336..759K}, and the resulting
electrodynamics equations, 
 while simpler to deal with as now one only needs to consider the behavior of electromagnetic
fields constrained by the force-free condition, still represent a non-linear coupled system of partial differential equations.
The electrodynamics equations are then augmented by those describing the dynamical behavior of the spacetime and the neutron-star matter, thus
complex and time-consuming simulations are typically needed to study the system's evolution. Nevertheless,
for the scenario of interest here -- i.e. the late stages of a magnetized binary merger -- and for the purpose of our work, we can make
use of a hybrid approach, combining the formalism of Ref.~\cite{Palenzuela:2013hsa} described above (whereby the neutron stars are treated as point-like objects
satisfying ordinary differential equations of motion that incorporate the relevant gravitational and scalarization effects) together
with a \textit{unipolar model} to account for magnetospheric effects. Both these approaches are supported by 
simulations of the complete problem in the context of neutron star mergers (for ST theories and non-magnetized systems~\cite{Palenzuela:2013hsa}) and
binary neutron star and black hole-neutron star mergers in the context of magnetosphere interactions 
in GR~\cite{Ponce:2014sza,Palenzuela:2013kra,Palenzuela:2013hu,Paschalidis:2013jsa}. As argued
in the previous section, a reliable model accounting for magnetosphere interactions within GR
should also be applicable to the case of ST theories of gravity. In what follows, we therefore describe the main aspects of
the unipolar model.

\subsubsection*{Magnetosphere interactions and luminosity}
\label{sec:magnetosphereInteractions}

A useful model to estimate the electromagnetic energy radiated by a
magnetized neutron star binary is based on the \textit{unipolar inductor}~\cite{1969ApJ...156...59G,Hansen:2000am}.
Such a model has recently been further analyzed for neutron star binaries in Ref.~\cite{Lai:2012qe} and confronted with
fully dynamical simulations including plasma effects, finding good agreement in the
obtained luminosity~\cite{Palenzuela:2013kra,Palenzuela:2013hu,Ponce:2014sza}. Similar conclusions 
have been obtained in the model's application to black hole-neutron star binaries~\cite{2011ApJ...742...90M,Paschalidis:2013jsa}. 
We next describe briefly the main ingredients required by this model for our purposes.

We assume that both stars are magnetized, their magnetic fields are
dipolar and are non-spinning\footnote{Spins introduce only minor modifications --as tidal locking can not occur--
and thus will not affect the conclusions of this work.}. Further, we assume that one star has a larger magnetization
than its companion, and study the electromagnetic radiation due to the interaction
of their magnetospheres. As the orbit tightens, as described
in section \ref{subsec:STdynamics}, ST effects will cause deviations from the GR orbital behavior, which induce a stronger Poynting flux. 
It is important to stress here that for realistic magnetic field strengths,
electromagnetic effects do not backreact on the orbital evolution of the binary~\cite{Ioka:2000yb,Anderson:2008zp}.
Under such assumptions, one can estimate the luminosity of the binary~\cite{Lai:2012qe} as:
\begin{multline}
	\label{eq:lums-vrel_sepn}
	\mathcal{L} \approx 10^{38}	\left(\frac{v_{rel}}{c}\right)^2
				 	 \left(\frac{B_*}{10^{11}{\rm G}}\right)^2
					 \left(\frac{R_*}{10{\rm km}}\right)^6\\\times \left(\frac{R_{\rm eff}}{10{\rm km}}\right)^2
					 \left(\frac{r}{100{\rm km}}\right)^{-6}
					{\rm erg/s}
\end{multline}
where $v_{rel}$ is the relative velocity of the binary (obtained with the PN expansion
as discussed in~\cite{Palenzuela:2013hsa}),
$B_*$ and $R_*$ are the magnetic field and the radius of the primary star,
and $r$ is the separation between the stars.
Induction takes place on the secondary star at a radius $R_{\rm eff}$, which is equal
to the star's radius $R_c$ when the secondary is unmagnetized. Otherwise,
$R_{\rm eff}$ is larger, as the {secondary's field} shields a region around it. We account
for this effect by defining (see Refs.~\cite{2012PhRvD..86j4035L,Palenzuela:2013hu,Palenzuela:2013kra}),
\begin{equation}
	\label{eq:effRadius}
	R_{\rm eff} = \max \left( r\left(\frac{B_c}{B_*}\right)^{1/3} , R_c\right),
\end{equation}
{where $B_c$ is the magnetic field of the secondary}.
Naturally, this effect is relevant at large separations, while for separations
 $r \leq R_c \left(\frac{B_c}{B_*}\right)^{-1/3}$ the effective radius reduces to the star's radius $R_c$.
Within GR calculations of the quasi-adiabatic regime of binary neutron star systems (i.e. at large separations), 
estimates have been obtained from Eq.~(\ref{eq:lums-vrel_sepn}) by replacing $v_{rel}$ with its Keplerian
expression and $R_{\rm eff}=R_c$, giving rise to a dependence $\mathcal{L} \simeq r^{-7}$ 
(e.g. Refs.~\cite{Hansen:2000am,Lai:2012qe}). When the shielding effect given by Eq.~(\ref{eq:effRadius})
is considered, the luminosity follows a softer dependence $\mathcal{L} \simeq r^{-5}$
~\cite{Palenzuela:2013kra,Ponce:2014sza}. However, in ST theories, deviations from
Keplerian motion are possible, and we thus employ both Eqs.~(\ref{eq:lums-vrel_sepn}) and (\ref{eq:effRadius}) 
in our calculations. 

It is important at this point to stress two limitations of our model.
First, some mechanism must act to convert the obtained Poynting flux to observable radiation, and this will
involve some conversion efficiency. This work does not address this issue, but rather focuses on
estimating the Poynting flux alone. The fact that the system that we study shares many common features with
pulsars, where observable radiation in multiple bands is observed, gives us some degree of confidence that
some energy conversion to observable radiation will take place~\cite{Palenzuela:2013kra}.  
Second, this work concentrates on possible emissions prior to the merger. As the merger takes
place, the two stars will become tidally disrupted, merge into a hypermasive neutron star and
possibly form an accreting black hole. This rich dynamics will naturally have strong associated
luminosities, which are not accounted for here, as we concentrate on the pre-merger stage.

\section{Results}
\label{sec:Results}

\subsection{Quasi-circular case}
We first study binary systems in quasi-circular (i.e. zero eccentricity) orbits and
consider four different sets of masses. These configurations are chosen  so that
they undergo at least one of the key scalarization processes
described in section~\ref{subsec:STdynamics}, {while
still being consistent with available solar-system and binary-pulsar data.}
More specifically, the configurations we consider are:
\begin{itemize}
\item case {\bf LE}, with low- and equal-mass stars ($M_1 = M_2 = 1.41M_\odot$), 
which undergo dynamical scalarization but do not produce dipolar radiation.
\item case {\bf HE}, with high- and equal-masses stars ($M_1 = M_2 = 1.74M_\odot$),
which undergo spontaneous scalarization for $\tilde{\beta}=-4.5$ and dynamical scalarization for $\tilde{\beta}=-4.2$,
but do not produce dipolar radiation.
\item case {\bf LU}, with low- and unequal-mass stars ($M_1 = 1.41 M_\odot$, $M_2 = 1.64 M_\odot$), which
undergo dynamical as well as induced scalarization, and produce dipolar radiation.
\item case {\bf HU}, with high- and unequal-mass stars ($M_1 = 1.52 M_\odot$, $M_2 = 1.74 M_\odot$), which 
undergo dynamical and induced scalarization (in the lower-mass star) and
spontaneous scalarization (in the higher-mass star), and produce dipolar radiation.
\end{itemize}

Additionally, we consider different possible magnetizations of
each star, and examine the characteristics of the resulting electromagnetic luminosity.
Henceforth, we will refer to the primary/companion star as the more/less massive one and, in the case of
equal mass configurations, as the more/less strongly magnetized one.
To adopt realistic configurations, we recall that 
the standard formation channel of neutron-star binaries indicates that
the most likely configurations involve a magnetically
dominant (primary) star with a significantly less magnetized companion
(secondary)~\cite{2001ApJ...557..958C,lrr-2008-8,lrr-2006-6}.
To explore a range of possible options, we consider three ratios of the magnetizations between the stars,
namely $b \equiv B_c/B_* = (0.1, 0.01, 0.001)$. 
Also, for simplicity we assume that the stars' magnetic dipoles are aligned with the orbital angular momentum. 
Notice that
this is not a restrictive assumption,
as it yields reasonably good estimates for the expected power in more general configurations~\cite{Ponce:2014sza}.
Finally, we restrict our analysis to the ST theories
that yield the largest  differences in the binary dynamics, while satisfying existing experimental constraints, i.e.
we take the coupling parameter of the ST theory to be $\tilde{\beta} = -4.5$  or $\tilde{\beta}=-4.2$ (so as to satisfy binary pulsar constraints
\footnote{{Note however, as already mentioned,
that the viable range for $\beta$ depends slightly on the equation of state adopted for the neutron stars.}}), and we adopt
a small value for the asymptotic value of the scalar field $\varphi_0=10^{-5}$ to pass solar system tests.
 For comparison purposes, we also include the 
corresponding GR results. 
The list of the cases considered, as well as a summary of the main
results (e.g., the total radiated electromagnetic energy for each case),
is given in Table~\ref{table:cases_results}.

\begin{table*}[!]
\begin{tabular}{c||c|c|c|c	|c}
	\hline
		&	Masses	&	Magnetic field	&		&	\multicolumn{2}{c}{Total $E_{rad}$ [erg]} 	\\
	Case	&(in $M_\odot$) &	ratio, $b \equiv B_c/B_*$ &	Theory	&	at 35 km	& at 30 km	\\
	\hline\hline
	low equal-mass 
		& $M_1 = 1.41 $	&	$0.1$	&	
		&	1.74 $\times10^{41}$
		&	(1.90,1.93,1.93)$\times10^{41}$
	\\
	(\bf{LE})	& $M_2 = 1.41 $	&	$0.01$	&	$\tilde{\beta}=(-4.5, -4.2$); GR
		&	4.02$\times10^{40}$
		&	(4.72,4.83,4.83)$\times10^{40}$
	\\
		&	&	$0.001$	&		
                &	2.31$\times10^{40}$
		&	(3.02,3.12,3.12)$\times10^{40}$
	\\
	\hline
	high equal-mass
		& $M_1 = 1.74 $	&	$ 0.1$	&	
                &       (1.27,1.49,1.55)$\times10^{41}$
                &       (1.44,1.66,1.78)$\times10^{41}$
	\\
	(\bf{HE})	&  $M_2 = 1.74$	&	$ 0.01$	&	$\tilde{\beta}=(-4.5, -4.2)$; GR
                &       (3.06,3.61,3.87)$\times10^{40}$
		&	(4.24,4.78,5.42)$\times10^{40}$
	\\
		&	&	$ 0.001$	&	
                &       (1.74,2.19,2.45)$\times10^{40}$
		&	(2.92,3.37,4.01)$\times10^{40}$
	\\
	\hline
        low unequal-mass
                &	$M_1 = 1.41 $        &       $ 0.1$ &     
                &       $(1.18,1.29,1.41)\times10^{41}$
		&	$(1.27,1.49,1.55)\times10^{41}$
        \\
        (\bf{LU})        &	$M_2 = 1.64 $	&       $ 0.01$        &       $\tilde{\beta}=(-4.5, -4.2)$; GR
                &       $(2.65,3.16,3.24)\times10^{40}$
		&       $(3.07,3.61,3.87)\times10^{40}$
        \\
                &       &       $0.001$       &      
                &       $(1.32,1.74,1.83)\times10^{40}$
		&       $(1.73,2.19,2.45)\times10^{40}$
        \\
	\hline
        high unequal-mass
                &	$M_1 = 1.52 $	&       $ 0.1$ &      
		&       $(0.913,1.05,1.18)\times10^{41}$
                &       $(0.996,1.13,1.3)\times10^{41}$
        \\
        (\bf{HU})        &	$M_2 = 1.74 $	&       $0.01$        &       $\tilde{\beta}=(-4.5, -4.2)$; GR
		&       $(2.07,2.37,2.7)\times10^{40}$
                &       $(2.45,2.75,3.22)\times10^{40}$
        \\
                &       &       $ 0.001$       &     
		&       $(1.06,1.2,1.5)\times10^{40}$
                &       $(1.43,1.58,2.02)\times10^{40}$
        \\
	\hline\hline
\end{tabular}
\caption{Quasi-circular cases studied in this work.
	The last two columns show the total electromagnetic energy radiated ($E_{rad} = \int \mathcal{L} dt$)
	for binaries starting at an initial separation of 180 km apart,
	until a separation of 35 and 30 km respectively.
}
\label{table:cases_results}
\end{table*}

\begin{figure*}[!]
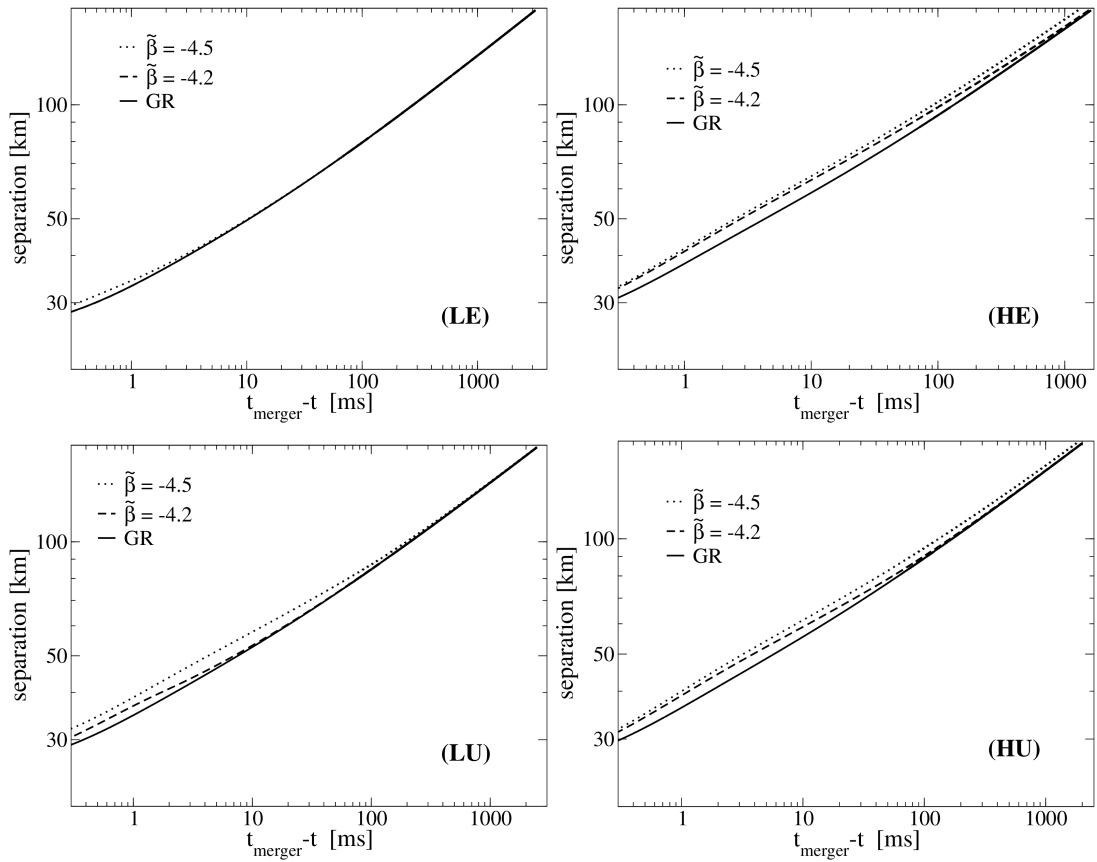

        \includegraphics[width=0.4\textwidth]{{{./sepn-time_14-14-loglog}}}
        \includegraphics[width=0.4\textwidth]{{{./sepn-time_174-174-loglog}}}
        \\
        \vspace{2mm}
        \includegraphics[width=0.4\textwidth]{{{./sepn-time_141-164-loglog}}}
        \includegraphics[width=0.4\textwidth]{{{./sepn-time_152-174-loglog}}}
	\caption{
		Separation shown as a function of $t_{\rm{merger}} -t $ in log-scale,
		being $t_{\rm{merger}}$ the time at which
		the stars come into contact (i.e. the separation equals $R_*+R_c$).
		The panels display: the \textbf{LE} (top-left panel), \textbf{HE} (top-right panel),
		\textbf{LU} (bottom-left panel) and  \textbf{HU} (bottom-right panel) binaries,
		for $\tilde{\beta}=-4.5$ (dotted lines), $-4.2$ (dashed lines) and GR (solid lines).
		Recall that magnetic field effects on the binary's motion are negligible,
		thus the differences in the trajectories are solely due to the underlying gravity theory.
		In the \textbf{LE} binary, the trajectories for the three cases are almost identical, as
		there is almost no scalarization until very late in the inspiral.
		{These examples show that at any given separation, the time to merger is shorter (or equal) in ST 
theories than in GR.}
		}
       \label{fig:sepn-time}
\end{figure*}

\begin{figure*}
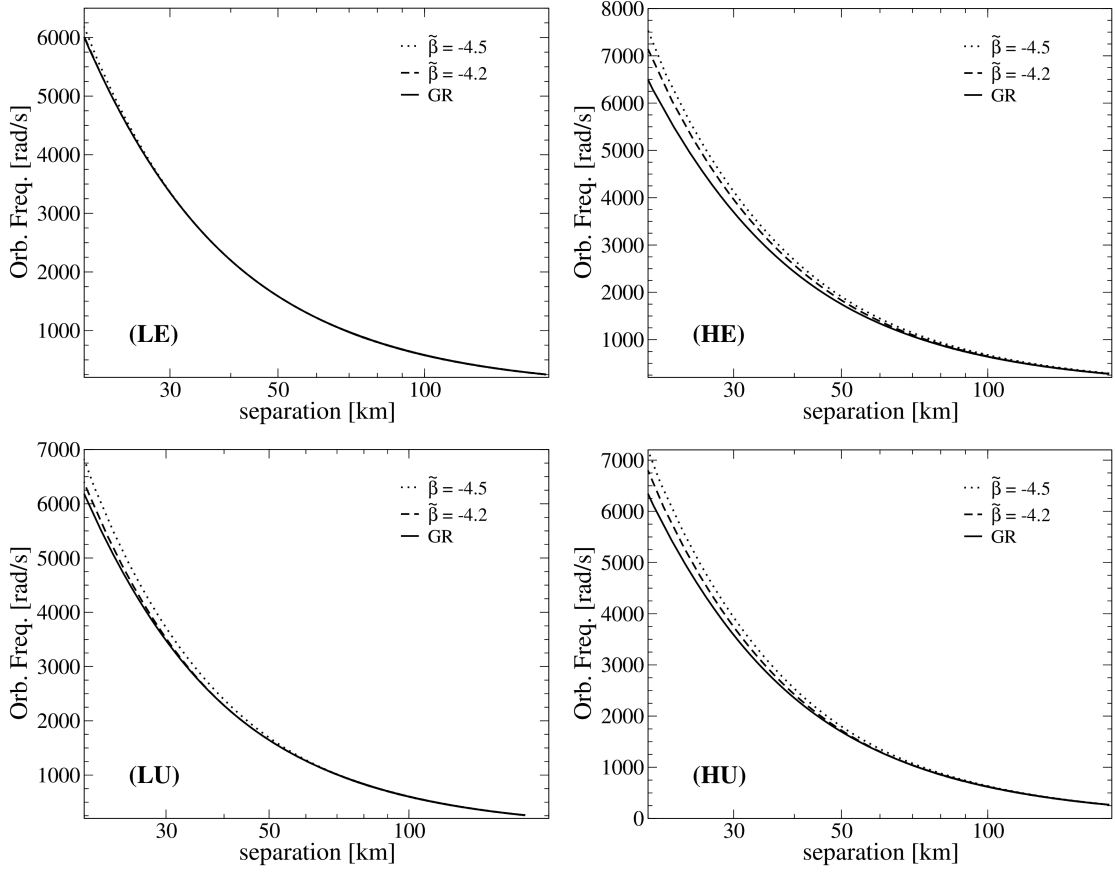

        \includegraphics[width=0.4\textwidth]{{{./thetadot-sepn_14-14}}}
	~
        \includegraphics[width=0.4\textwidth]{{{./thetadot-sepn_174-174}}}
        \\
        \vspace{2mm}
        \includegraphics[width=0.4\textwidth]{{{./thetadot-sepn_141-164}}}
	~
        \includegraphics[width=0.4\textwidth]{{{./thetadot-sepn_152-174}}}

	\caption{Orbital frequency as a function of separation (in log-scale), for the same parameters shown in Fig.~\ref{fig:sepn-time}.
		}
	\label{fig:thetadot-sepn}
\end{figure*}

As mentioned above, the binary dynamics in the ST theories that we study can show clear departures from the GR behavior. 
In particular, the binary's orbital frequency/separation can increase faster than in GR 
for high neutron-star masses and low values of $\tilde{\beta}$, because of the enhanced gravitational attraction due to scalar effects
and the possible dipolar emission of scalar waves. { Such behavior is illustrated in  
Fig.~\ref{fig:sepn-time}, which shows the separation for the four binaries considered, in
 ST theories with $\tilde{\beta} =-4.5$ and  $\tilde{\beta}=-4.2$, as well as in GR, as a function of
the time remaining until the merger. (The merger is defined
as the time at which the separation equals $R_*+R_c$). As illustrated in this figure,  
in the ST theories that we consider the time to merger from a given separation is 
always either shorter or equal than in GR. As we discuss next, in the former case differences in the electromagnetic flux of energy arise.}
The complementary Fig.~\ref{fig:thetadot-sepn} shows how all binary evolutions cover approximately the same frequency range,
although for higher masses and lower values of $\tilde{\beta}$, any given frequency is achieved at a larger separation.

\begin{figure*}[!]
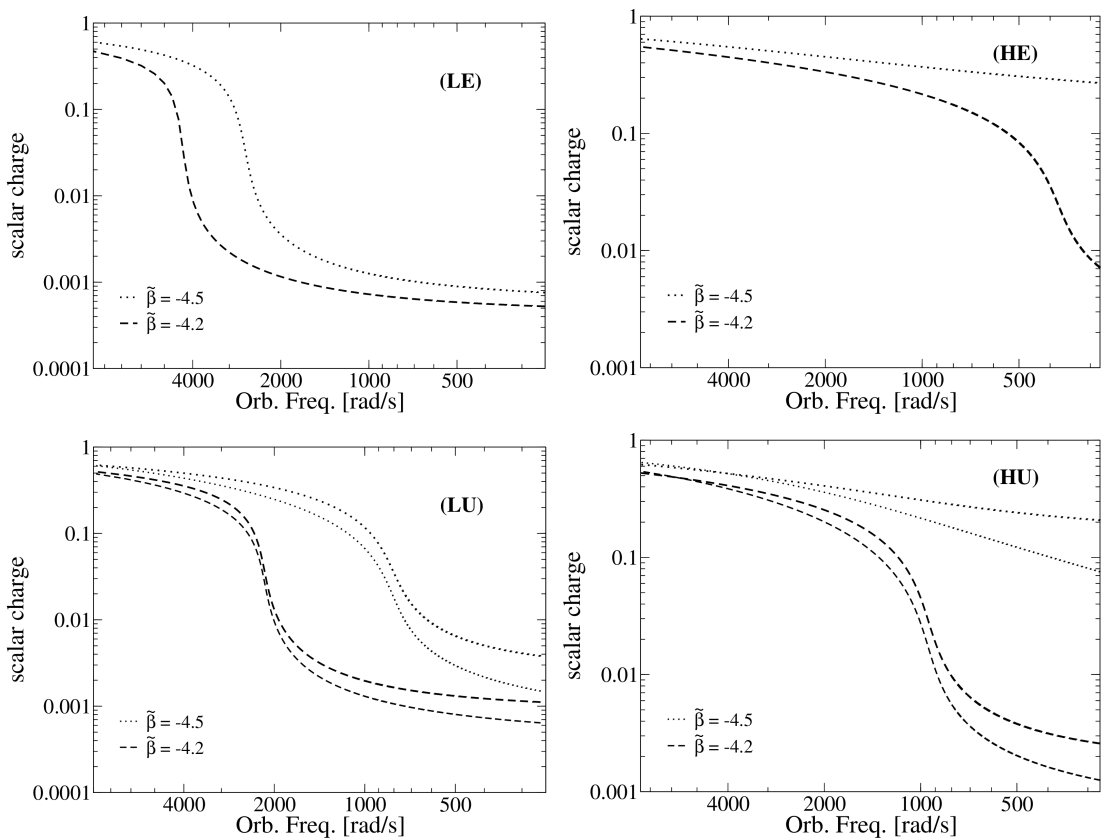

        \includegraphics[width=0.4\textwidth]{{{./ScChg-thetadot_14-14}}}
        ~\includegraphics[width=0.4\textwidth]{{{./ScChg-thetadot_174-174}}}
        \\
        \vspace{2mm}
        \includegraphics[width=0.4\textwidth]{{{./ScChg-thetadot_141-164}}}
        ~\includegraphics[width=0.4\textwidth]{{{./ScChg-thetadot_152-174}}}
	\caption{Scalar charge for the same binaries as in  Fig.~\ref{fig:sepn-time},
		as a function of the orbital frequency,
		for ${\tilde{\beta}}=-4.5$ (dotted lines) and $-4.2$ (dashed lines).
		The top panels correspond to equal-mass binaries, and both stars have
		the same scalar charge for a given value of ${\tilde{\beta}}$. 
		On the other hand, the bottom panels correspond to unequal-mass binaries, and
		the scalar charges are different.
		In these cases in particular, the upper line with the same style
		(i.e. the same value of ${\tilde{\beta}}$) refers to the more massive star.
		}
	\label{fig:ScChg-sepn_thetadot}
\end{figure*}

As discussed, these effects are caused
by the scalar charges (or equivalently the sensitivities) that each star
acquires as the orbits tightens. The behavior of the scalar charge with respect to the orbital frequency for
each case considered is shown in Fig.~\ref{fig:ScChg-sepn_thetadot}, 
where the orbital frequency is determined instantaneously (i.e. as the derivative  of the
azimuthal coordinate).
From this figure, it is clear that scalarization effects are quite weak in the
\textbf{LE} case, because the scalar charges remain negligible for most of the evolution and only rise
to $\alpha\sim {\cal O}(1)$ at relatively short separations (i.e. high frequencies).
In particular, the scalar charges remain $\lesssim 10^{-3}$ essentially until
the final plunge toward merger. 
This behavior is in contrast with the remaining cases, where scalarization effects
are considerably stronger at the earlier stages, and therefore have a clear impact on
the dynamics, especially for $\tilde{\beta}=-4.5$.
Such behavior comes about because in the {\bf HE}, {\bf LU} and {\bf HU} cases the more massive star either spontaneously scalarizes
already in isolation, 
which induces a time-dependent running of the scalar charge of the companion (induced scalarization), or both stars scalarize dynamically in the late inspiral
(after which the scalar charges further grow by mutual induced scalarization).  As mentioned, as the scalar charges grow,
the gravitational attraction between the stars gets stronger than in GR, and for unequal-mass binaries the system
also emits dipolar radiation, clearly affecting the binary's dynamics.

\begin{figure*}[!]
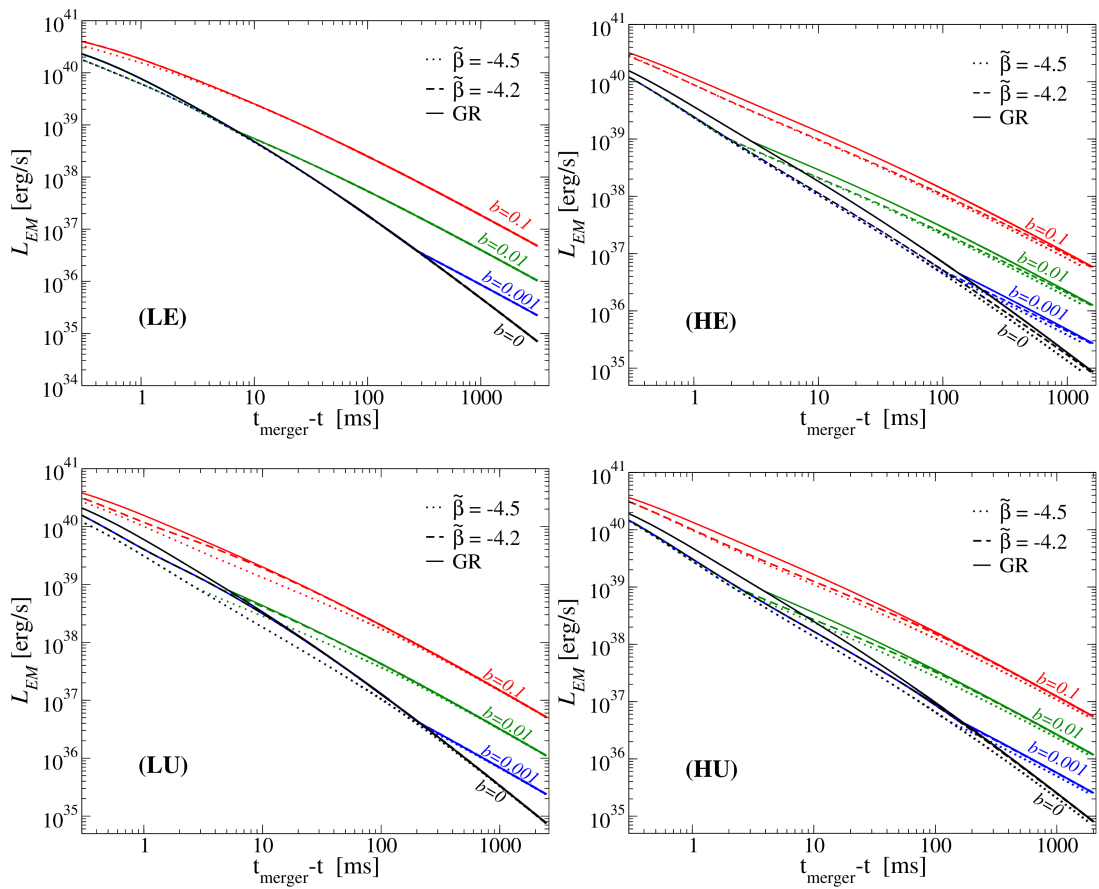

        \includegraphics[width=0.4\textwidth]{{{./beta_-4.5-4.2-GR--1.41-1.41--time}}}
        \includegraphics[width=0.4\textwidth]{{{./beta_-4.5-4.2-GR--1.74-1.74--time}}}
        \\
        \vspace{2mm}
        \includegraphics[width=0.4\textwidth]{{{./beta_-4.5-4.2-GR--1.41-1.64--time}}}
        \includegraphics[width=0.4\textwidth]{{{./beta_-4.5-4.2-GR--1.52-1.74--time}}}
	\caption{Luminosity vs. time to merger, for the same parameters shown in Fig.~\ref{fig:sepn-time}.
                Different lines in each panel indicate the \textit{ratio} between
		the secondary and primary magnetic fields (from top to bottom):
		$b=0.1$ (in red), $b=0.01$ (in green), $b=0.001$ (in blue)
		and a non-magnetized secondary (i.e. $b=0$, in black).
		}
	\label{fig:lum-time}
\end{figure*}

\begin{figure*}[!]
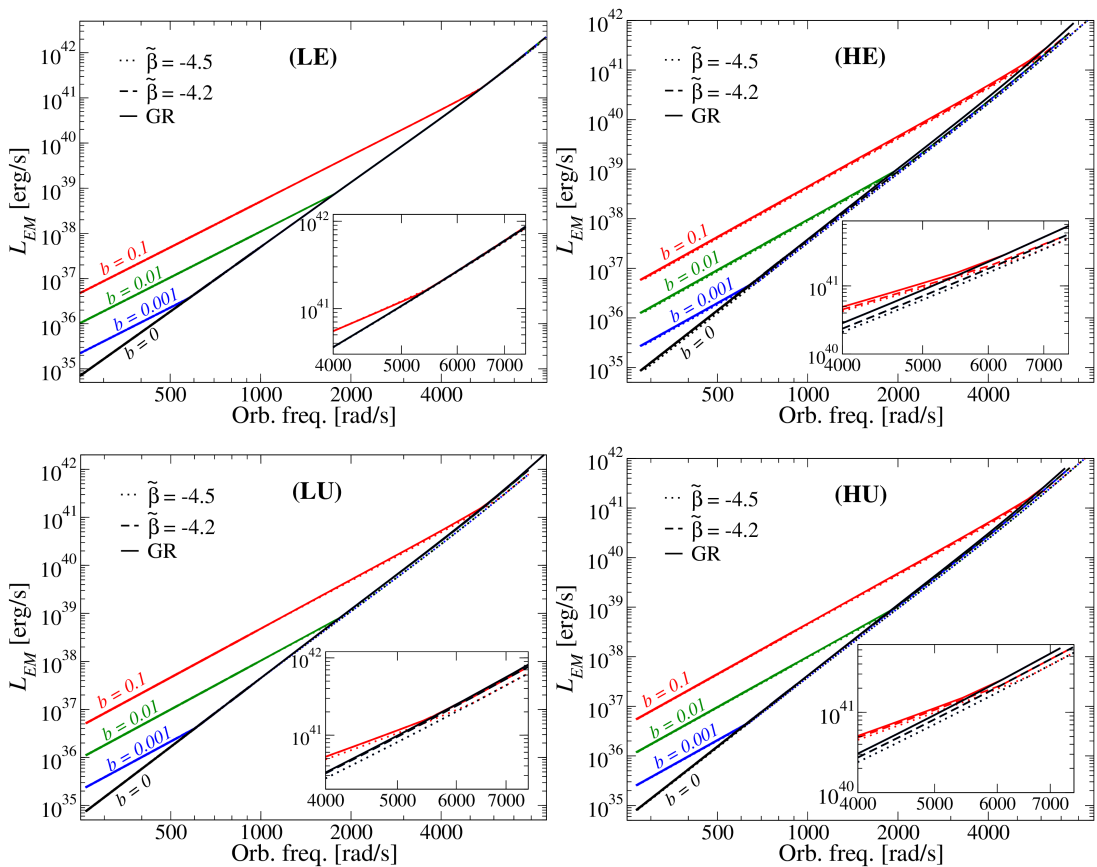

        \includegraphics[width=0.4\textwidth]{{{./beta_-4.5-4.2-GR--1.41-1.41--freq_linInset}}}
        \includegraphics[width=0.4\textwidth]{{{./beta_-4.5-4.2-GR--1.74-1.74--freq_linInset}}}
        \\
        \vspace{2mm}
        \includegraphics[width=0.4\textwidth]{{{./beta_-4.5-4.2-GR--1.41-1.64--freq_linInset}}}
        \includegraphics[width=0.4\textwidth]{{{./beta_-4.5-4.2-GR--1.52-1.74--freq_linInset}}}
        \caption{Luminosity as a function of orbital frequency, for the same parameters shown in Fig.~\ref{fig:lum-time}.
		}
        \label{fig:lum-freq}
\end{figure*}

The dynamical behavior that we have just described has direct consequences on the Poynting flux produced by
the system as the stars' magnetospheres interact. As described in Sec.~\ref{sec:magnetospherePlasma},
we estimate such flux via the enhanced unipolar inductor model, which depends on the orbital evolution as well
as on the magnetization of the binary components. The results for the luminosity
are displayed in Figs.~\ref{fig:lum-time} and~\ref{fig:lum-freq} as a function
of time and orbital frequency. 
At a broad level, since all cases cover similar frequency ranges and separations -- from $\approx 180$ km to merger for the cases considered -- 
all the obtained luminosities are comparable and of the order of $10^{39}$--$10^{41} \mbox{erg/s}$
(the precise value depending on the binary's mass and the magnetic field ratio between the stars)
for a primary with magnetic field of $10^{11}\mbox{G}$.

\begin{figure*}[!]
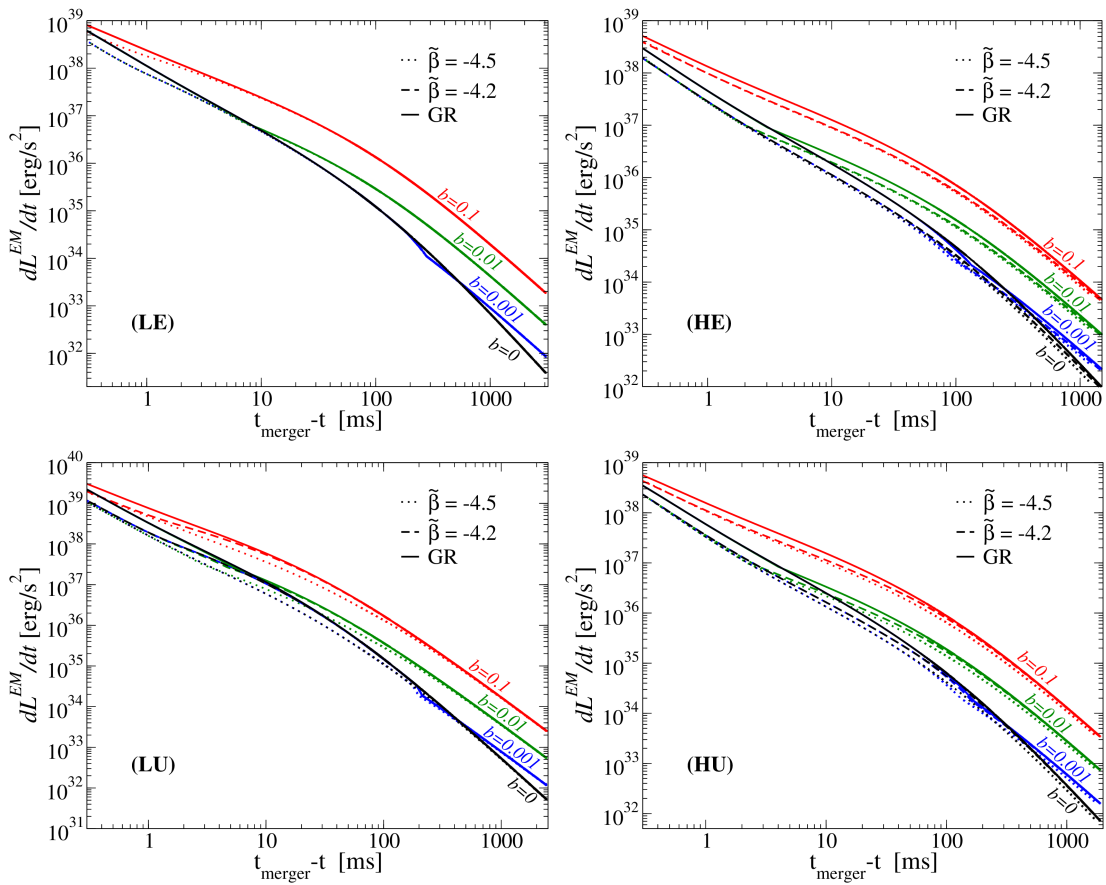

        \includegraphics[width=0.4\textwidth]{{{./dLdt_avg--beta_-4.5-4.2-GR--1.41-1.41--time}}}
        ~\includegraphics[width=0.4\textwidth]{{{./dLdt_avg--beta_-4.5-4.2-GR--1.74-1.74--time}}}
	\\
	\vspace{1mm}
	\includegraphics[width=0.4\textwidth]{{{./dLdt_avg--beta_-4.5-4.2-GR--1.41-1.64--time}}}
	~\includegraphics[width=0.4\textwidth]{{{./dLdt_avg--beta_-4.5-4.2-GR--1.52-1.74--time}}}
	\caption{Time derivative of the luminosity as a function of time to merger,
		 for the same parameters shown in Fig.~\ref{fig:lum-time}. 
		Note that the differences are less significative for the low-mass binaries.
		}
	\label{fig:dLdt-time}
\end{figure*}

\begin{figure*}
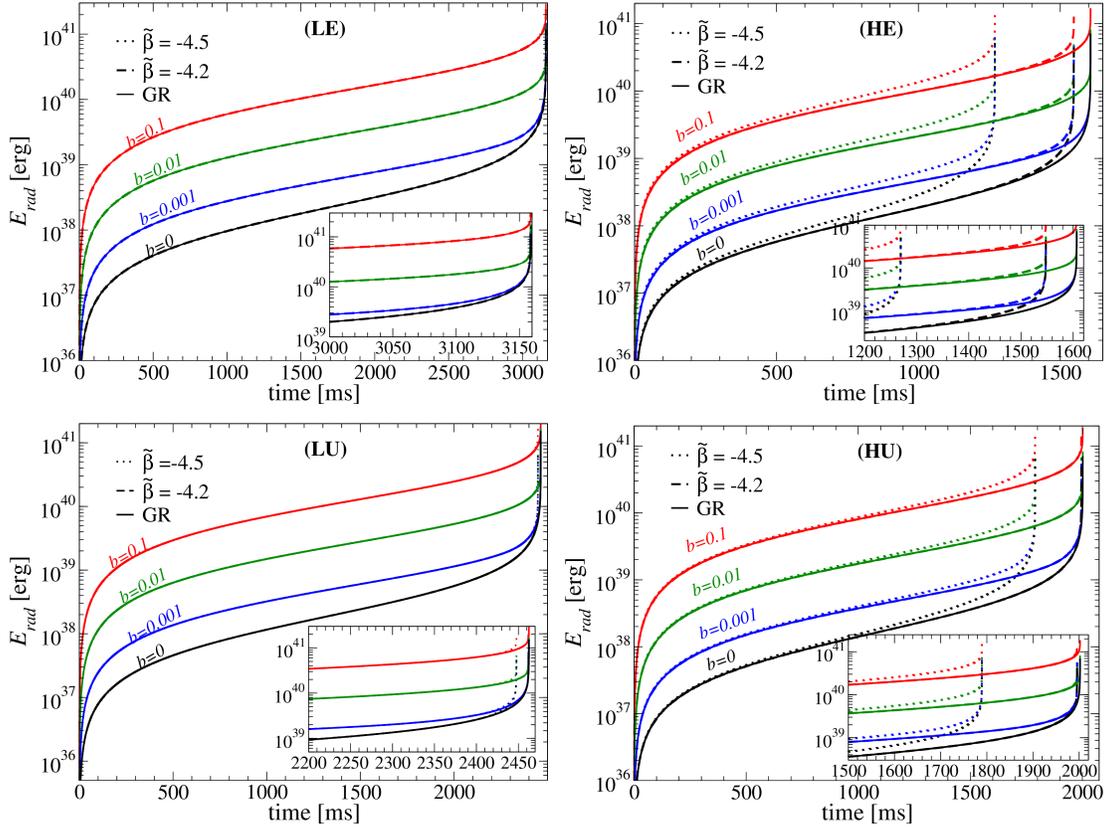

	\includegraphics[width=0.4\textwidth]{{{./intLum--beta_-4.5-4.2-GR--1.41-1.41--time_wINset}}}
        ~\includegraphics[width=0.4\textwidth]{{{./intLum--beta_-4.5-4.2-GR--1.74-1.74--time_wINset}}}
	\\
	\vspace{2mm}
	\includegraphics[width=0.4\textwidth]{{{./intLum--beta_-4.5-4.2-GR--1.41-1.64--time_wINset}}}
	~\includegraphics[width=0.4\textwidth]{{{./intLum--beta_-4.5-4.2-GR--1.52-1.74--time_wINset}}}
	\caption{
		 Total energy radiated in electromagnetic waves (i.e. the time integral of the luminosity)
 as a function of time elapsed from an initial separation of approximately 180 km.
		for the same parameters shown in Fig.~\ref{fig:lum-time}.
		Insets show zoom-ins of the total energy radiated at late times, when the binary is close to the merger.
		}
	\label{fig:Erad-time}
\end{figure*}

A closer inspection, e.g.~of Fig.~\ref{fig:lum-time}, shows that ST effects are evident for the more massive
binaries and for lower values of $\tilde{\beta}$, especially near the merger.
This behavior is also visible
in the luminosity rate of change with time, as illustrated in Fig.~\ref{fig:dLdt-time}. As can be seen,
for higher masses and lower values of $\tilde{\beta}$, the dynamics proceeds at a faster pace in the ST theories under consideration than in GR,
thus inducing a less powerful electromagnetic flux {at a given time to merger (since for a given $t_{\rm merger}-t$, the binary's separation is larger
in the ST theories that we consider than in GR, c.f. Fig.~\ref{fig:sepn-time})}. This conclusion is further supported by
the {total radiated electromagnetic energy, as a function of the time elapsed from an initial separation of 180 km (Fig.~\ref{fig:Erad-time}). }
As time progresses, binaries governed by the ST theories show a clear departure  from the general-relativistic behavior.

\begin{figure*}[!]
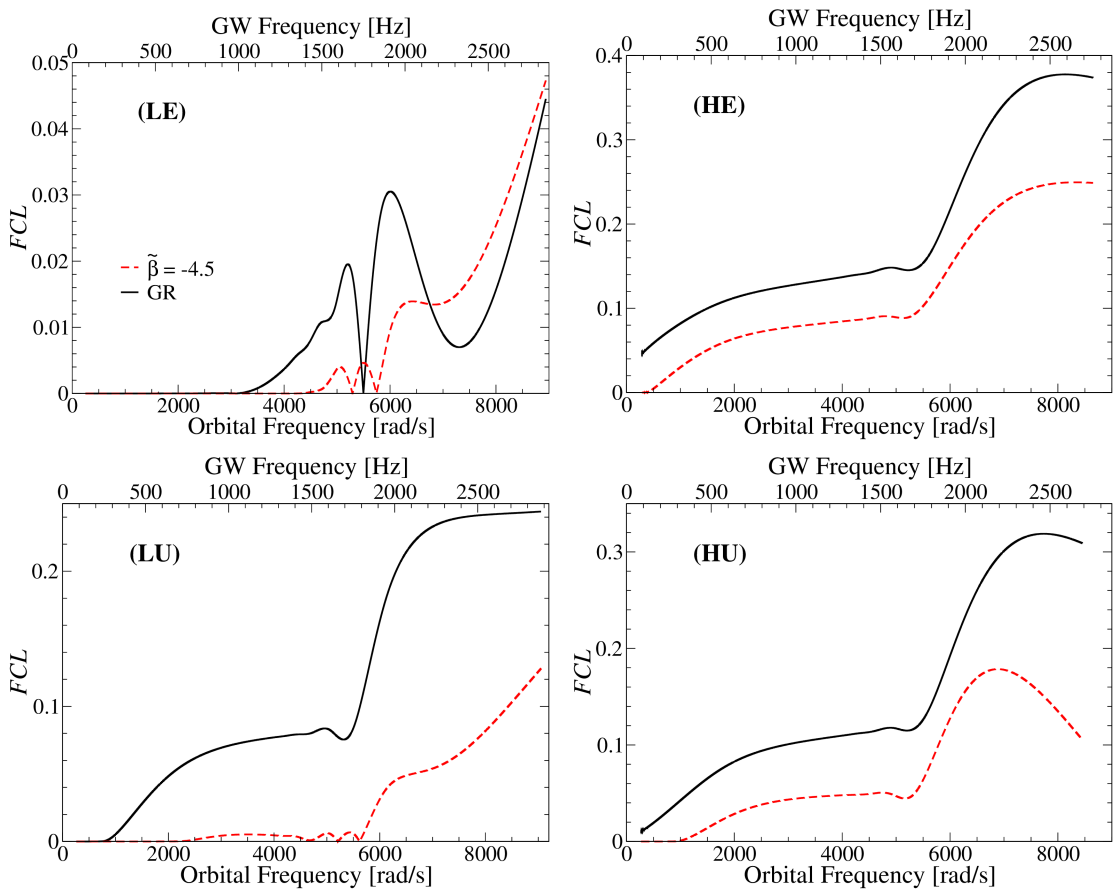

        \includegraphics[width=0.4\textwidth]{{{./FrcChg_LE}}}
        ~
        \includegraphics[width=0.4\textwidth]{{{./FrcChg_HE}}}
        \\
        \vspace{2mm}
        \includegraphics[width=0.4\textwidth]{{{./FrcChg_LU}}}
        ~
        \includegraphics[width=0.4\textwidth]{{{./FrcChg_HU}}}

        \caption{\textit{Fractional change} of the luminosity, 
		$FCL \equiv {|\mathcal{L}_{\rm ST}-\mathcal{L}_{\rm GR}|}/{\mathcal{L}_{\rm GR}}$,
		for the \textbf{LE} (top-left panel), \textbf{HE} (top-right panel),
                \textbf{LU} (bottom-left panel) and  \textbf{HU} (bottom-right panel) binaries
                as a function of orbital and GW frequencies (assuming  $b = 0.1$ for concreteness and
                clarity's sake).
		The solid black line shows the $FCL$ of a ST with $\tilde{\beta}=-4.5$ with respect to GR,
		while the dashed red line represents the $FCL$ of a ST with $\tilde{\beta}=-4.2$ (again with respect to GR). 
               }
        \label{fig:FracChg_lum-freq}
\end{figure*}

To further illustrate the differences in luminosities obtained, Fig.~\ref{fig:FracChg_lum-freq}
presents the \textit{fractional change in luminosity},
$FCL \equiv {|\mathcal{L}_{\rm ST}-\mathcal{L}_{\rm GR}|}/{\mathcal{L}_{\rm GR}}$. (For concreteness we adopt $b=0.1$, as
 all other values of $b$ display a similar behavior.)
As can be seen, the \textbf{LE} case displays at most a $5\%$ difference, while the other ones
(\textbf{LU}, \textbf{HE} and \textbf{HU})  show departures of up to 40\% away from the GR prediction. Importantly, the relative differences display a frequency
dependent behavior, growing strongly toward higher frequencies.

\subsection{Eccentric binaries}
We now turn our attention to binaries with non-negligible eccentricity. Such configurations
are not expected to represent a large fraction of the sources detectable by advanced gravitational-wave detectors,
as gravitational emission tends to circularize binaries by the time they enter the detector's sensitive band.
Nevertheless, as already illustrated e.g.~in Ref.~\cite{Healy:2009zm,Gold:2011df,East:2012ww}, eccentric binaries provide
an excellent laboratory to test diverse and extreme physics. As we show here, this is also
the case for possible electromagnetic counterparts driven by magnetosphere interactions in non-GR gravity theories. To illustrate
this point, we here consider the following two cases:
an equal-mass binary with low masses $M_1 = M_2 = 1.52 M_\odot$,
and an unequal-mass binary with $M_1 =1.52 M_\odot$ and $M_2 = 1.74 M_\odot$.
Following Ref.~\cite{East:2012ww}, we study the dynamics and the emitted electromagnetic flux,
in two different configurations:
``mild-eccentricity'' orbits, where the initial apastron/periastron are at separations of 
about 220 km/80 km respectively (i.e. $e \sim 0.47$);
and,
``high-eccentricity'' orbits, where the initial apastron/periastron are at separations of 
about 220 km/50 km respectively (i.e. $e \sim 0.63$).

\begin{figure*}
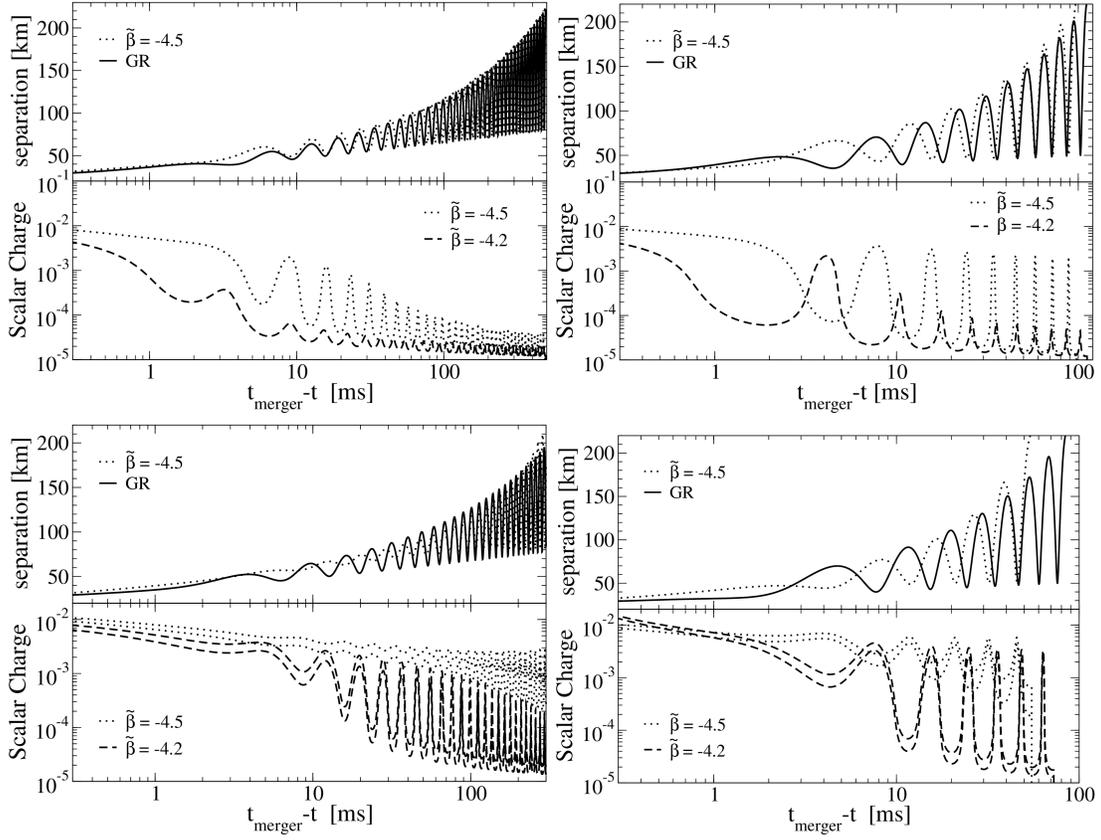

        \includegraphics[width=0.4\textwidth]{{{./sepnScChg-time_151-151_lowEcc}}}
        \includegraphics[width=0.4\textwidth]{{{./sepnScChg-time_151-151_hghEcc}}}
        \\
        \vspace{1mm}
        \includegraphics[width=0.4\textwidth]{{{./sepnScChg-time_152-174_lowEcc}}}
        \includegraphics[width=0.4\textwidth]{{{./sepnScChg-time_152-174_hghEcc}}}
	\caption{Eccentric binaries: evolution of the separation for a binary governed
by GR and ST with $\tilde \beta=-4.5$ and scalar charges of the binary's components (for binaries
governed by ST with $\tilde \beta=-4.5$ and $\tilde \beta=-4.2$ for comparison purposes)
		as a function of time to merger, for an equal-mass binary (top panels) and
		an unequal-mass binary (bottom panels), for low-eccentricity orbits (left column) and
		high-eccentricity orbits (right column) [see text for the details on the exact masses and eccentricities].
As the orbit progresses, and scalar charges become significant, the eccentricity is reduced more rapidly in the ST cases
as illustrated by the increasingly reduced differences between local maxima and minima of the binary's separation.
		}
	\label{fig:sepn_eccs}
\end{figure*}

\begin{figure*}
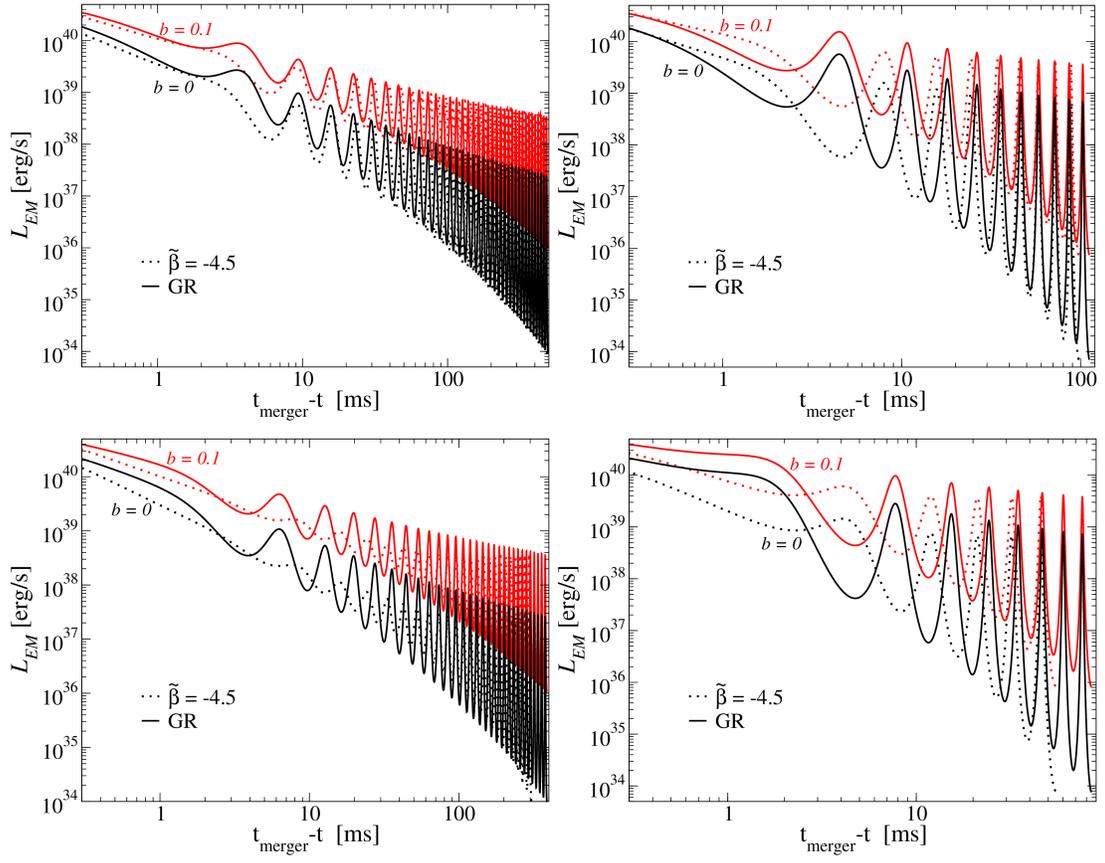

        \includegraphics[width=0.4\textwidth]{{{./beta_-4.5-4.2-GR--1.51-1.51_lowEcc-b_0-0.1}}}
        \includegraphics[width=0.4\textwidth]{{{./beta_-4.5-4.2-GR--1.51-1.51_hghEcc-b_0-0.1}}}
        \\
        \vspace{2mm}
        \includegraphics[width=0.4\textwidth]{{{./beta_-4.5-4.2-GR--1.52-1.74_lowEcc-b_0-0.1}}}
        \includegraphics[width=0.4\textwidth]{{{./beta_-4.5-4.2-GR--1.52-1.74_hghEcc-b_0-0.1}}}
        \caption{Luminosity vs time to merger, for the same parameters shown in Fig.~\ref{fig:sepn_eccs}.
		For clarity's sake we show here the two extremal magnetic ratios considered,
		$b = 0.1$ and $b = 0$ (i.e. a non-magnetized companion).
		}
        \label{fig:lums_eccs}
\end{figure*}

These eccentric binaries reveal another interesting phenomenon allowed in the ST theories under study, namely
a sequence of successive ``scalarization/descalarization'' cycles~\cite{Palenzuela:2013hsa}. These
produce a strong modulation of the scalar charges (see Fig.~\ref{fig:sepn_eccs}),
with a consequent impact on the dynamics, and possibly observable signatures both in the gravitational
and electromagnetic signals. For instance, the binary's orbit circularizes more rapidly for smaller values of
$\tilde{\beta}$ as illustrated in Fig.~\ref{fig:sepn_eccs}. As can also be seen in that figure, except for
the low-, equal-mass binary with mild eccentricity (which shows negligible differences between GR and ST theories), all the other cases display increasingly
marked deviations from GR as $\tilde{\beta}$ decreases and higher masses are considered. This behavior
has a direct impact on the electromagnetic flux, as shown in Fig.~\ref{fig:lums_eccs}.
Of particular interest is the fact that the orbital eccentricity induces, in all cases, an oscillatory
behavior in the flux intensity with a frequency and amplitude modulated by a growing trend as the orbit shrinks. 
{Interestingly, the growth rate is different in the cases governed by ST theories for sufficiently low values of $\tilde{\beta}$.}
Note { however} that misalignment of the stars' dipole moments induces oscillations in the resulting luminosity
even in the quasi-circular case~\cite{Ponce:2014sza}. 

\begin{figure*}
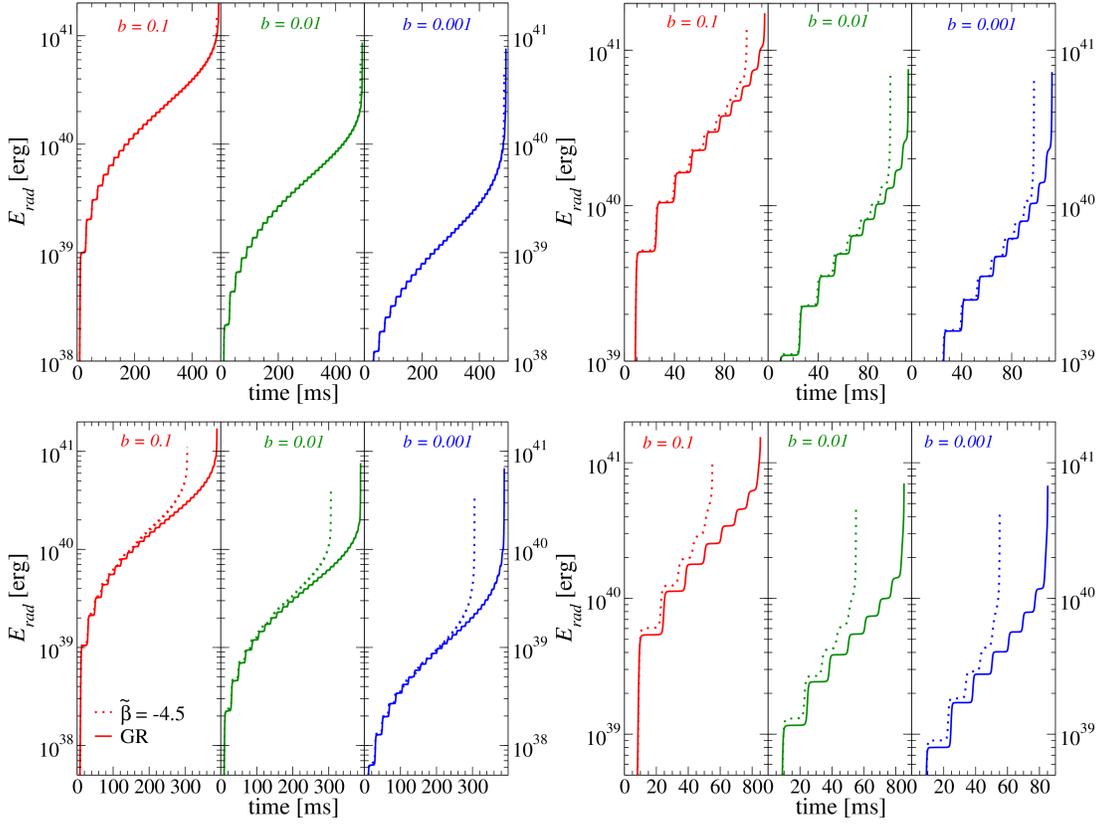

        \includegraphics[width=0.4\textwidth]{{{./intLum--beta_-4.5-4.2-GR--1.51-1.51_lowEcc_x3}}}
        \includegraphics[width=0.4\textwidth]{{{./intLum--beta_-4.5-4.2-GR--1.51-1.51_hghEcc_x3}}}
	\\
	\vspace{2mm}
        \includegraphics[width=0.4\textwidth]{{{./intLum--beta_-4.5-4.2-GR--1.52-1.74_lowEcc_x3}}}
        \includegraphics[width=0.4\textwidth]{{{./intLum--beta_-4.5-4.2-GR--1.52-1.74_hghEcc_x3}}}

        \caption{Energy radiated by eccentric binaries,
as a function of the time elapsed from an initial binary separation of approximately 220 km,
		for the same parameters shown in Fig.~\ref{fig:sepn_eccs}.
		We show, in contiguous panels, three different magnetic ratios (from left to right within each plot):
		$b = 0.1$ (in red), $0.01$ (in green), and $0.001$ (in blue).
		Note that in the case of low-eccentricity orbits (left column), the radiated energy has a smoother (continuous) trend,
		while in the case of high-eccentricity orbits (right column) the radiated energy is more jagged;
		this is a consequence of the oscillatory features in the luminosity, and ultimately the effect of the different dynamics of each binary.
		}
        \label{fig:intLum_eccs}
\end{figure*}

{Finally, we note that the more rapid orbital evolution of strongly scalarized binaries 
can cause  the same total energy to be radiated within a significantly shorter time. This is shown
in Fig.~\ref{fig:intLum_eccs}, where we stress that we are showing the total radiated energy as a function
of time (and \textit{not} time to merger).
}

\section{Discussion}
\label{sec:Discussion}
The results shown in this work indicate that, within the context of electromagnetic emission
induced by magnetosphere interactions, binary neutron star systems could produce signals with
clear deviations from the GR expectation, depending on the gravity theory.
Our results open up the possibility of exploiting compact binary systems
to test gravitational theories by combining electromagnetic and gravitational signals. For instance, in the particular case of the ST theories of gravity considered here,
deviations in the gravitational-wave signals away from the GR prediction could be detected for binaries that undergo scalarization sufficiently
early in the advanced LIGO band~\cite{Sampson:2014qqa,Taniguchi:2014fqa} (naturally, how early in frequency
depends strongly on the signal-to-noise ratio of the signal). Such binaries, as illustrated here, also show distinctive departures
in the Poynting flux luminosity emitted from the system, when compared to the behavior within GR. The fact that deviations
could be observed with advanced LIGO if they take place at sufficiently low frequencies in the gravitational-wave signal 
is a consequence of the detector's sensitivity curve (i.e. sufficiently early scalarization is needed to build up
enough signal-to-noise ratio in band). For this reason, scalarization effects
at high frequencies in the ST theories that we consider would be difficult to detect by gravitational-wave signals alone, unless the detector
is tuned for higher sensitivity in the higher frequency band. Without such tuning, however, complementary electromagnetic
signals could be exploited. As we illustrated here, such an opportunity does indeed appear to be possible. 

In particular, our results indicate deviations in the expected electromagnetic luminosity in ST theories for {sufficiently low} 
values of the coupling $\tilde{\beta}<0$. This observation
 is not enough (in itself) to assess the nature of the underlying gravitational theory, 
unless a good estimate of the stars' magnetization is available.
However, we find that the luminosity's strength and -- most importantly -- rate of change with time
exhibit differences {from the GR behavior} in the ST theories that we study, and this could provide precious information that
can be used to test the gravity theory. Indeed, in many respects, the rate of change of the luminosity would provide
analog information to the ``chirping'' gravitational signal, though in this case not limited by advanced LIGO's/VIRGO's noise
curve, which in its canonical configuration raises sharply  before the kHz frequency at which the merger takes place. 

As discussed in Ref.~\cite{Palenzuela:2013kra}, the Poynting flux from binary systems can indeed induce high-energy signals from the system, which can be exploited for this goal. Naturally, an important question to consider
is the distance at which these sources could be detected in the electromagnetic band. Such a distance
estimate must take into account both the type of expected radiation from the system, as well as the typical sensitivities of 
the various observational facilities. A rough estimate can be computed by using the peak luminosities
$\simeq 10^{40} (B/10^{11}{\rm G})^2$ erg/s prior to the merger, and assuming that a relativistically expanding
electron-positron wind (sourced by energy dissipation and magnetohydrodynamical waves in between the stars)
produces an x-ray signal~\cite{Hansen:2000am}, preceding or coincident with the merger.
These signals may be detectable with ISS-Lobster, by virtue of its high sensitivity and wide field of view,
and may be seen up to distances of $\sim (B/10^{11}G)$~Mpc, if one assumes  a fiducial 10\% efficient conversion of Poynting flux.
However, it is important to stress that the advent of gravitational-wave {detectors} would  
 potentially allow one to increase the accessible distances. Focused efforts by both
the gravitational-wave and electromagnetic-signal communities are ongoing to make the most of these ``multimessenger'' opportunities.

\vskip0.3cm
\textit{Acknowledgments:} 
We thank N. Cornish, L. Sampson and N. Yunes for interesting discussions.
EB acknowledges support from
the European Union's Seventh Framework Programme (FP7/PEOPLE-2011-CIG)
through the  Marie Curie Career Integration Grant GALFORMBHS PCIG11-GA-2012-321608.
LL acknowledges support by NSERC through a Discovery Grant and CIFAR. LL thanks
the Institut d'Astrophysique de Paris and the ILP LABEX (ANR-10-LABX-63),
for hospitality during a visit supported through the Investissements d'Avenir program under
reference ANR-11-IDEX-0004-02.
Research at Perimeter Institute is supported through Industry Canada
and by the Province of Ontario through the Ministry of Research and Innovation.

\normalem

\bibliography{./STpnBNSem}

\end{document}